\documentclass[fleqn,usenatbib]{mnras}
\usepackage{amsmath}
\usepackage{xspace}
\usepackage{mathptmx}
\usepackage{txfonts}
\usepackage{xcolor}

\usepackage[T1]{fontenc}
\usepackage{ae,aecompl}

\usepackage{graphicx}	
\usepackage{amssymb}	
\usepackage{booktabs}
\usepackage{wasysym}

\let\oldhref\href
\renewcommand{\href}[2]{\oldhref{#1}{\hbox{#2}}}

\definecolor{colorl1}{RGB}{0, 51, 153}
\definecolor{colorl2}{RGB}{153, 0, 0}
\definecolor{colorl3}{RGB}{179, 179, 0}
\definecolor{colorl4}{RGB}{51, 102, 0}

\definecolor{colorw1}{RGB}{51, 102, 255}
\definecolor{colorw2}{RGB}{255, 51, 0}
\definecolor{colorw3}{RGB}{255, 214, 51}
\definecolor{colorw4}{RGB}{51, 204, 51}

\newcommand{\hMpc}{{\ifmmode{h^{-1}{\rm Mpc}}\else{$h^{-1}$Mpc}\fi}}
\newcommand{\Mpc}{{\ifmmode{{\rm Mpc}}\else{Mpc}\fi}}
\newcommand{\hkpc}{{\ifmmode{h^{-1}{\rm kpc}}\else{$h^{-1}$kpc}\fi}}
\newcommand{\kpc}{{\ifmmode{ {\rm kpc}}\else{{\rm kpc}}\fi}}
\newcommand{\kms}{{\ifmmode{ {\rm km\,s^{-1}}}\else{ ${\rm km\,s^{-1}}$}\fi}}
\newcommand{\hMsun}{{\ifmmode{h^{-1}{\rm {M_{\astrosun}}}}\else{$h^{-1}{\rm{M_{\astrosun}}}$}\fi}}
\newcommand{\Msun}{{\ifmmode{{\rm M}_{\astrosun}}\else{${\rm M}_{\astrosun}$}\fi}}
                        
\newcommand{\Mhalo}{{\ifmmode{M_{\rm halo}}\else{$M_{\rm halo}$}\fi}}
\newcommand{\Rvir}{{\ifmmode{R_{\rm vir}}\else{$R_{\rm vir}$}\fi}}
\newcommand{\Mvir}{{\ifmmode{M_{\rm vir}}\else{$M_{\rm vir}$}\fi}}
\newcommand{\Mstar}{{\ifmmode{M_{\rm star}}\else{$M_{\rm star}$}\fi}}
\newcommand{\Vrot}{{\ifmmode{V_{\rm rot}}\else{$V_{\rm rot}$}\fi}}
\newcommand{\ltsima}{$\; \buildrel < \over \sim \;$}
\newcommand{\gtsima}{$\; \buildrel > \over \sim \;$}
\newcommand{\lsim}{\lower.5ex\hbox{\ltsima}}
\newcommand{\gsim}{\lower.5ex\hbox{\gtsima}}

\def\lesssim{\mathrel{\hbox{\rlap{\hbox{\lower4pt\hbox{$\sim$}}}\hbox{$<$}}}}
\def\gtrsim{\mathrel{\hbox{\rlap{\hbox{\lower4pt\hbox{$\sim$}}}\hbox{$>$}}}}

\newcommand{\Fig}[1]{Fig.~\ref{#1}}
\newcommand{\beq}{\begin{equation}}
\newcommand{\eeq}{\end{equation}}
\def\beqa{\begin{eqnarray}}
\def\eeqa{\end{eqnarray}}
\def\LCDM{\ensuremath{\Lambda}CDM}

\def\head{ \vbox to 0pt{\vss \hbox to 0pt{\hskip 440pt\rm
      LA-UR-10-07069\hss} \vskip 25pt}}

\def \kms {\ifmmode  \,\rm km\,s^{-1} \else $\,\rm km\,s^{-1}$ \fi }
\def \kpc {\ifmmode  {\,\rm kpc}  \else ${\rm  kpc}$ \fi  }  
\def \hkpc {\ifmmode  {h^{-1}\rm kpc}  \else ${h^{-1}\rm kpc}$ \fi  }  
\def \hMpc {\ifmmode  {h^{-1}\rm Mpc}  \else ${h^{-1}\rm Mpc}$ \fi  }  
\def \Mpch {\ifmmode  {h^{-1}\rm Mpc}  \else ${h^{-1}\rm Mpc}$ \fi  }  
\def \Msun {\ifmmode {\rm M}_{\astrosun} \else ${\rm M}_{\astrosun}$ \fi} 
\def \hMsun {\ifmmode h^{-1}\,\rm M_{\astrosun} \else $h^{-1}\,\rm M_{\astrosun}$ \fi}
\def \Gyr {\ifmmode\, \rm Gyr \else $\,$Gyr \fi}

\def \LCDM {\ifmmode \Lambda{\rm CDM} \else $\Lambda{\rm CDM}$ \fi}
\def \sig8 {\ifmmode \sigma_8 \else $\sigma_8$ \fi} 
\def \OmegaM {\ifmmode \Omega_{\rm m} \else $\Omega_{\rm m}$ \fi} 
\def \Omegab {\ifmmode \Omega_{\rm b} \else $\Omega_{\rm b}$ \fi} 
\def \OmegaL {\ifmmode \Omega_{\rm \Lambda} \else $\Omega_{\rm \Lambda}$\fi} 
\def \Deltavir {\ifmmode \Delta_{\rm vir} \else $\Delta_{\rm vir}$ \fi}
\def \rhocrit {\ifmmode \rho_{\rm crit} \else $\rho_{\rm crit}$ \fi}
\def \rhou {\ifmmode \rho_{\rm u} \else $\rho_{\rm u}$ \fi}
\def \zc {\ifmmode z_{\rm c} \else $z_{\rm c}$ \fi}

\def\Mdyn {\ensuremath {M_{\textrm{dyn}}(<r_{23.5})}~}
\def\Mstar {\ensuremath {M_{*}(<r_{23.5})}~}
\def\r23_5 {\ensuremath {r_{23.5}}~}

\title[Abundance Matching tested on Small Scales] {Abundance matching tested on small scales with galaxy dynamics}

\author[A.V. Macci\`o et al.]{Andrea V. Macci\`o$^{1,2,3}$\thanks{E-mail: maccio@nyu.edu},
St\'ephane Courteau$^4$, Nathalie N.-Q. Ouellette$^{4,5}$,
\newauthor{Aaron A. Dutton$^{1}$}
\\
$^{1}$New York University Abu Dhabi, PO Box 129188, Abu Dhabi, United Arab Emirates \\
$^2$Center for Astro, Particle and Planetary Physics (CAP$^3$), New York University Abu Dhabi\\
$^3$Max-Planck-Institut f\"ur Astronomie, K\"onigstuhl 17, 69117 Heidelberg, Germany \\
$^4$Queen's University, Department of Physics, Engineering Physics and Astronomy, Kingston, Ontario, Canada\\
$^5$Universit\'e de Montr\'eal, D\'epartement de physique, C.P. 6128, Succ. Centre-ville, Montr\'eal, Qu\'ebec, Canada
}

\date{Accepted XXX. Received YYY; in original form ZZZ}

\pubyear{2020}

\begin{document}

\label{firstpage}
\pagerange{\pageref{firstpage}--\pageref{lastpage}}
\maketitle
\begin{abstract}

We present a comprehensive test of the relation between stellar and total mass in galaxies
as predicted by popular models based on abundance matching (AM) techniques.
We use the ``Spectroscopy and H-band Imaging of Virgo cluster galaxies'' (SHIVir) survey
with photometric and dynamical profiles for 190 Virgo cluster galaxies 
to establish a relation between the stellar and dynamical masses measured within the isophotal radius $r_{23.5}$. 
Various dark matter and galaxy scaling relations are combined with results from the NIHAO suite of hydrodynamical simulations to recast AM predictions in terms of these observed quantities.  Our results
are quite insensitive to the exact choice of dark matter profile and halo response to baryon collapse.
We find that theoretical models reproduce the slope and normalization of the observed
stellar-to-halo mass relation (SHMR) over more than three orders of magnitude in stellar
mass $(10^8 < M_*/\Msun < 2 \times 10^{11})$.
However, the scatter of the observed SHMR exceeds that of AM predictions by a factor $\sim$5.
For systems with stellar masses exceeding $5 \times 10^{10}~\Msun$, AM overpredicts the
observed stellar masses for a given dynamical mass.  The latter offset may support
previous indications of a different stellar initial mass function in these massive galaxies. 
Overall our results support the validity of AM predictions on a wide dynamical range.

\end{abstract}

\begin{keywords}
cosmology: theory -- dark matter -- galaxies: formation -- galaxies: kinematics and dynamics -- methods: numerical
\end{keywords}

\section{Introduction}\label{sec:introduction}

In a universe dominated by dark matter and dark energy, galaxy formation is a complex mix
of hierarchical mass assembly, gas cooling, star formation  and secular evolution.

In recent years, the scaling relation between stellar mass and halo (or total) mass has emerged as a fundamental trend to be reproduced by any model, theoretical
or numerical, aiming to explain and characterize the intricate process of galaxy formation.
The Stellar-to-Halo Mass Relation (SHMR hereafter), pionered by \cite{moster2010} and subsequently refined
by other authors \citep[e.g][]{Behroozi2013, Moster2013, Kravtsov2016}, has become
one of the premier tests for galaxy formation models at low and high redshifts  \citep[e.g.][]{Stinson2013, Hopkins2014, Schaye2015, Wang2015}.

While numerous galaxy scaling relations emerge directly from the direct combination
of fundamental observables, such as luminosity and velocity for the Faber-Jackson and
Tully-Fisher relations \citep{Faber1976, Tully1977} or size-velocity relations \citep{Courteau2007},
the SHMR is a hybrid mix of observations and theory, based on the Abundance Matching (AM)
paradigm which assumes that a galaxy with a given volume abundance and mass range (e.g. in number per cubic megaparsec per mass dex) will inhabit dark matter haloes with the same space density.

Despite its widespread appeal, the SHMR is not easily measured and calibrated
given the intrinsic challenge of directly measuring the total (halo) mass of galaxies.
A few modeling techniques like gravitational lensing and satellite dynamics enable
the measurement of a total mass in the outskirts of galaxies, but these methods
are typically only applied to (fairly) massive galaxies and such measurements are
rather limited \citep[e.g.][]{Courteau2014}. 

In this paper, we present a different approach to this problem: rather than exploiting
SHMRs constructed from observations alone, we combine a suite of observed galaxy
scaling relations with hydrodynamical cosmological simulations of galaxy formation 
\citep[namely the NIHAO project:][]{Wang2015} to recast Abundance Matching predictions in a more
practical framework. 

Our technique enables a comparison of theoretical predictions for the SHMR from
two of the most commonly used AM models by \cite{Behroozi2013} and \cite{Moster2013}
with observational results from the 
SHIVir survey of structural parameters for 190 spatially-resolved Virgo cluster galaxies \citep{Ouellette2017}.
Our comprehensive SHMR 
provides a superb opportunity to test various precepts inherent to theoretical galaxy formation models. 

This Letter is organized as follows: sections \ref{sec:observations} and \ref{sec:simulations} presents an outline of the SHIVir survey and 
of our numerical simulations respectively. Our results on the comparison between AM predictions and observational data
are shown in section \ref{sec:results}, while section \ref{sec:conclusions} contains our discussion and conclusions.

\section{Observational data} \label{sec:observations}

We take advantage of the 
``Spectroscopy and H-band Imaging of Virgo cluster galaxies" (SHIVir) survey  
of 190 Virgo cluster of galaxies to extract the structural parameters needed for
our study, such as stellar masses, dynamical masses, and isophotal sizes, among others. 

SHIVir is an optical and near-infrared survey combining analysis of Sloan Digital Sky
Survey (SDSS) photometry, deep H-band photometry and long-slit spectroscopy acquired
uniformly on multiple telescopes and instruments.  The SHIVir catalog of 190 Virgo
cluster galaxies (VCGs) covers all morphological types over a stellar mass range of
$10^{7.8}~\Msun$ to $10^{11.5}~\Msun$, including 133 VCGs that have resolved
kinematic profiles (rotation curves and dispersion profiles).
The SHIVir photometric catalog is presented in \citet{McDonald2009} [H-band imaging], \citet{Roediger2011a,Roediger2011b}
[multi-band imaging and stellar population analysis], 
and \citet[][and 2020 in preparation]{Ouellette2017} [long-slit spectroscopy].  We refer to
these papers for a detailed description of the catalog selection. The SHIVir photometric
data are available at http://www.astro.queensu.ca/virgo.  The SHIVir spectroscopic data
will be presented in Ouellette et al. (2020; in preparation).   Altogether, SHIVir is
a comprehensive database of photometric and kinematic parameters for Virgo cluster
galaxies with effective and isophotal grizH magnitudes and radii, rotational velocities,
velocity dispersions, stellar and dynamical masses.

Whenever possible, if light profiles reach to deep levels, structural parameters are
measured at R$_{23.5}$, the projected 2D isophotal radius corresponding to a surface brightness
of 23.5 mag sec$^{-2}$, typically measured in the reddest possible band \footnote{The i-band surface brightness at 23.5 mag sec$^{-2}$ corresponds roughly
to surface densities of 1-10 \Msun/pc$^2$.  }. 
In the rest of this Letter, we will use  R$_{23.5}$ to indicate the 2D isophotal radius and  r$_{23.5}$
to indicate its 3D counterpart;  all the SHMR parameters will be all measured within r$_{23.5}$, both in observations and simulations.

Stellar mass-to-light ratios, $M_*/L$, are recovered via global colours from the
spatially-resolved light profiles.  
Stellar masses, \Mstar, are inferred from suitably chosen stellar mass-to-light ratios
via transformations provided by \citet{Roediger2015}.

Dynamical masses, \Mdyn, for late-type systems assume spherical symmetry and are
computed as $\Mdyn = V_{rot,23.5}^2 \times \rm{r}_{23.5} / \rm{G}$, where $V_{rot,23.5}$ is
the average radial velocity of the gas measured at r$_{23.5}$.  $V_{rot,23.5}$ measurements
are corrected for inclination and redshift broadening \citep{Courteau2014,Ouellette2017}.
For early-type systems, \Mdyn, is computed using a transformation of the velocity dispersion, 
$\sigma$, measured at the effective radius, $r_e$, into circular velocity, $V_{rot,23.5}$ whereby: $V_{circ} = \sqrt {c}*\sigma_{r_e}$
and $c$ is a virial coefficient \citep[cf. Sec. 2.4][]{Ouellette2017}. 

Size errors are relatively small given their well-defined surface brightness reference, and the typical uniform uncertainty on $V_{rot,23.5}$ is 6 \kms.  Thefore, the typical uncertainty on $\Mdyn$ is estimated to be $\sim$10$-$15\%.  A conservative random uncertainty for stellar mass-to-light ratio, \Mstar, is 0.13 dex \citep{Stone2019}.
Systematic uncertainties (e.g. model-to-model differences) for the \Mstar, are however of order 0.3 dex \citep{Conroy2013a,Courteau2014,Roediger2015}.



\section{Simulations} \label{sec:simulations}

We take advantage of the cosmological hydrodynamical simulations NIHAO
(Numerical Investigation of Hundred Astrophysical Objects) for this project.
NIHAO is a large simulation suite of high resolution (zoom-in) simulations; for more details,
see \cite{Wang2015}.  The simulations are based on the gasoline2 code \citep{Wadsley2017},
and include metal cooling, chemical enrichment, star formation and feedback from supernovae (SN) 
and massive stars (the so-called Early Stellar Feedback). 
The cosmological parameters are set according to \citet{Planck2014}: Hubble parameter
$H_0$= 67.1 \kms Mpc$^{-1}$, matter density $\Omega_\mathrm{m}=0.3175$, dark energy density
$\Omega_{\Lambda}=1-\Omega_\mathrm{m} -\Omega_\mathrm{r}=0.6824$, baryon density
$\Omega_\mathrm{b}=0.0490$, normalization of the power spectrum $\sigma_8 = 0.8344$, 
slope of the inital power spectrum $n=0.9624$, and each galaxy is resolved with about
one million elements (dark matter, gas and stars).

The NIHAO simulations have been proven to be very successful in reproducing several observed
scaling relations like the SHMR \citep{Wang2015}, the disc gas mass and disc size relation \citep{Maccio2016}, 
the Tully-Fisher relation \citep{Dutton2017}, and the diversity of galaxy rotation curves \citep{Santos-Santos2018}.
NIHAO is indeed the needed tool for our goal of recasting the AM relation onto an observationally accessible plane.
This study uses all NIHAO simulated central galaxies (i.e. no satellites) with a stellar mass larger than $10^8$ M$_{\odot}$, for a total of ninety 
objects\footnote{Note that SHIVir galaxies are mainly satellites in the Virgo cluster, while NIHAO are central objects. 
Since we are interested in comparing masses with $r_{23.5}$ and given that the stellar body of a satellite galaxy is well protected against tidal effects
\citep{Smith2016}, we do not expect this difference to be significant.}

\section{Results}\label{sec:results}

\begin{figure}
\includegraphics[width=0.47\textwidth]{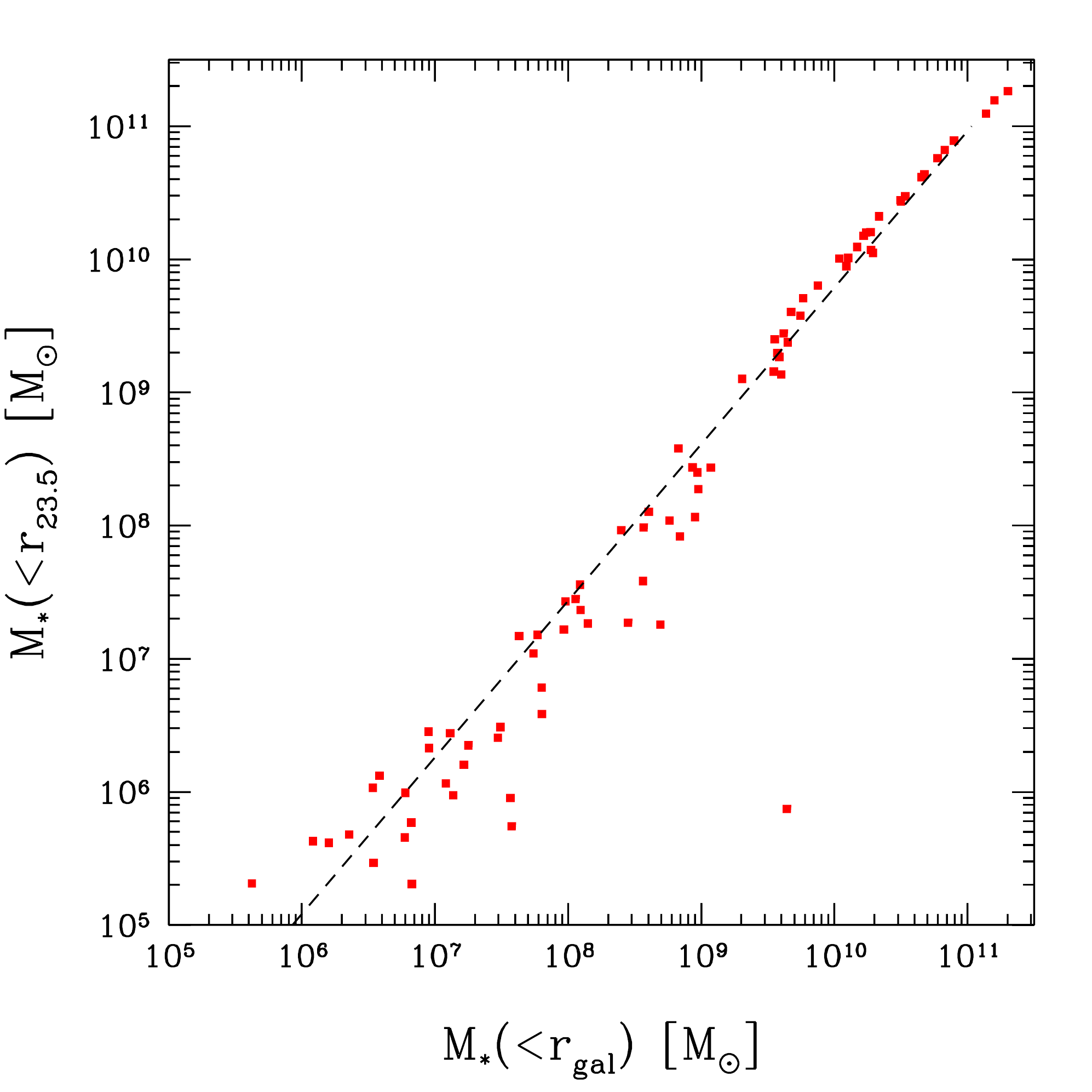}
\vspace{-.35cm}
\caption{Relation between the total stellar mass with the galactic radius $r_{gal}$, and the stellar mass within $r_{23.5}$ for the NIHAO galaxies.
The black dashed line is an orthogonal fit to the relation. The outlier in the bottom right corner is undergoing a merger and has been excluded 
from the fit.}
\label{fig:M23.5-Mtot}
\end{figure}

\begin{figure}
\includegraphics[width=0.47\textwidth]{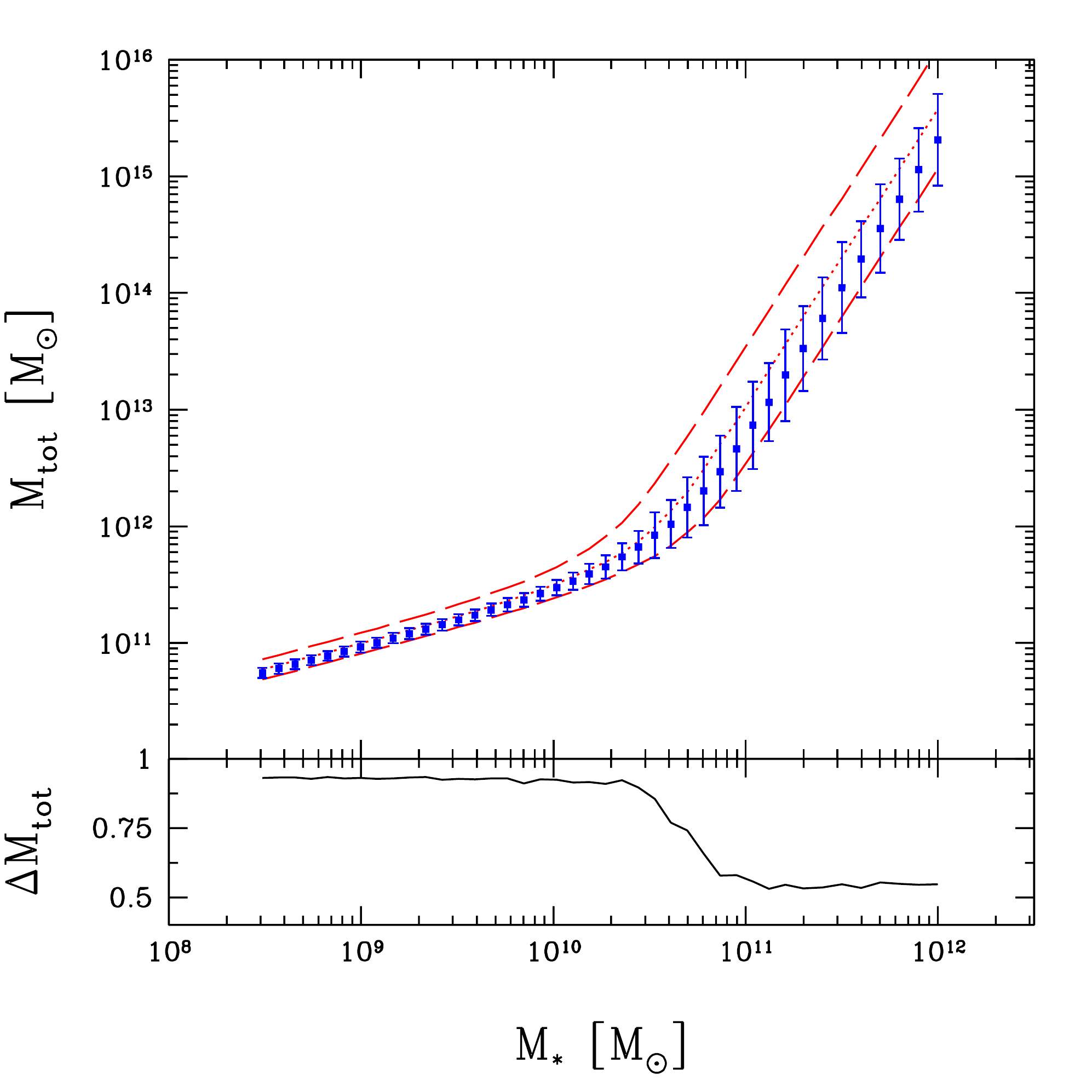}
\vspace{-.35cm}
\caption{The effects of scatter on halo mass at a given stellar mass on the inverted AM relation from Moster et al. (2013).
The dotted red line shows the relation obtained by a simple inversion (dashed lines represent the one-$\sigma$ scatter), while the blue points with error bars show the sample inverted relation when scatter is taken into account.
The lower panel shows the relative ratio between the two results.}
\label{fig:AMscatter}
\end{figure}
\begin{figure}
\includegraphics[width=0.47\textwidth]{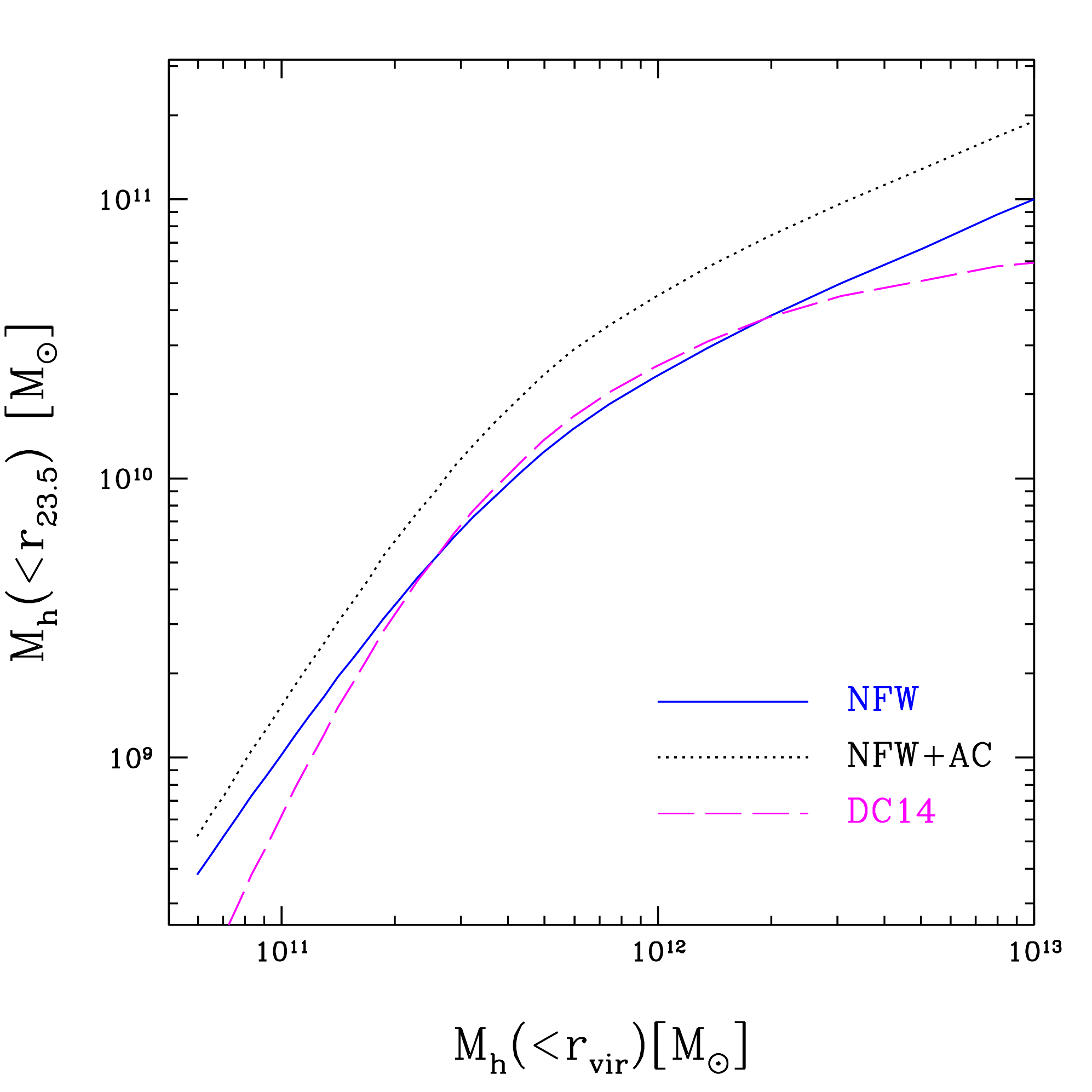}
\vspace{-.35cm}
\caption{Dark matter mass enclosed within \r23_5 as a function of total halo mass for the three different dark matter profiles adopted in our study.  Differences among these profiles are within a factor of three.}
\label{fig:Mh23.5}
\end{figure}

Our goal is to recast predictions based on AM techniques from the
$M_{*} - M_{\rm vir}$ plane into the observational motivated
\Mstar$- M_{\rm dyn}(<r_{23.5})$ plane. To this end, we must build a relation between $M_{*}$\footnote{We indicate the stellar mass of galaxy with $M_*(<r_{\rm gal})$, which is the  stellar mass within the galactic radius $r_{\rm gal}$ equal to 10\% of the virial radius \citep{Wang2015}.} and \Mstar and between $M_{\rm vir}$ and $M_{\rm dyn}(<r_{23.5})$.

To find the first relation we use the NIHAO simulations to compute for every galaxy the enclosed stellar mass $M_{*}(<r)$ as a function of $r$ and then verify for which radius $r$ this relation crosses the  relation between \r23_5 and \Mstar observed by \citet{Ouellette2017}:

\begin{equation} 
\log \left( \frac{ \r23_5 }{\rm kpc} \right) = -0.346 \times   \left( 7.51 - \log \left(  \frac{\Mstar}{\Msun} \right)\right).
\label{eq:one}
\end{equation}

The crossing point gives the values of $r_{23.5}$. With those in place, we can determine \Mstar for every NIHAO galaxy taking into account the  scatter of 0.13 dex in the observed relation (eq. \ref{eq:one}) \citep{Ouellette2017}.
We can now build a relation between \Mstar and  $M_*(<r_{\rm gal})$ based
on NIHAO. This relation, shown in Fig. \ref{fig:M23.5-Mtot}, is well represented by a simple linear fit: 
\begin{equation}
 \log \left(  \frac {\Mstar}  {\Msun}  \right) = 1.68 + 0.85 \times  \log \left( \frac{M_*(<r_{\rm gal})} {\Msun} \right).
\end{equation}
The fit has a median scatter of 0.3 dex, which we take into account when 
transforming  $M_*(<r_{\rm gal})$ into \Mstar. 
We must now derive the relation between the total mass of the galaxy within the virial radius and the dynamic (total) mass within \r23_5 .
The first quantity is obtained by inverting the AM relation, but the inversion is non-trivial due to the asymmetric scatter
in the inverted relation as well as the different halo abundances at low and high masses. In order to take these two effects into
account, we proceed as follows.  Assuming a constant scatter of 0.4 dex (at a fixed total mass) for each stellar mass \citep{Moster2018}, we
obtain (via inversion of the AM relation) minimum ($M_h^-$) and maximum ($M_h^+$) possible halo masses.  We then select
the actual halo mass in this interval with a probability distribution function $PDF\propto M_h^{-2}$ to account for the slope
of the halo mass function on these scales \citep[e.g.][]{Tinker2008} 
and thus, the different abundances of low and high mass haloes.

The results of this procedure are shown in \Fig{fig:AMscatter}. The blue points represent the inverted SHMR when the scatter is taken into account; 
these predict (at a fixed stellar mass) a lower value for the hosting dark matter halo mass than is expected from the direct inversion
(with no scatter) of the abundance matching result (red dotted line, the dashed line show the one-sigma scatter). 

To compute the total (dynamical) mass within \r23_5 , we must first infer the contribution of the dark matter component which can be obtained assuming a dark matter profile.  The density profile is then integrated from zero to the desired radius. 
The Navarro, Frenk \& White profile \citep[NFW][]{Navarro1996} is adopted, with a halo concentration from \cite{Dutton2014}.
Integration of the profile up to \r23_5 then yields $M_h(<r_{23.5})$.
Galaxy formation is known to alter the dark matter distribution, even though the precise effects are  still debated \citep[e.g.][]{Blumenthal1984,Gnedin2004, Read2005, Pontzen2012, DiCintio2014a, Chan2015}.
To isolate possible effects of galaxy formation, we consider two modes: 
i) a classical adiabatic contraction model based on \cite{Blumenthal1984}, and 
ii) the $\alpha-\beta-\gamma$ profile inferred by \citet[][DC14]{DiCintio2014b} based on the MaGICC simulations 
suite \citep{Maccio2012, Stinson2013} and subsequently verified with NIHAO simulations \citep{Tollet2016,Maccio2020}.
The results for the relation $M_h(<r_{23.5})$ for the three proposed profiles are shown in \Fig{fig:Mh23.5}.  
We will see in section \ref{sec:results} that the differences amongst the various mass profiles amount to less than 0.5 dex (factor 3)
and are thus insignificant relative to the scatter in the observed masses.

We can now compare the newly determined halo (dark matter) mass contribution to the total dynamical mass inside $r_{23.5}$,
with \Mdyn from our observations.  The latter is of course a combination of the three stellar, gas and dark matter components. 
First, we must correct the dark matter contribution resulting from pure N-body simulations 
(i.e. the NFW profiles and concentration mass relation), 
by multiplying our values of $M_h(<r_{23.5})$ by (1-$f_b$) where $f_b$ is the baryonic ratio: 
$f_b=\Omega_b/\Omega_m\approx 0.155$, for our choice of cosmological parameters (see section \ref{sec:simulations}).
The contribution of the stars and gas must also be accounted for.  The first one is given by \Mstar. 
The contribution from the gas is estimated from the scaling relation between total stellar mass and total (cold) gas mass from \citet[][with a scatter of a factor 2 around the mean]{Dutton2011}:
\begin{equation}
    \log \frac {M_g}{M_*} = -0.27 - 0.47 \log \left( \frac{M_*}{10^{10}\Msun} \right).
\end{equation}

The NIHAO simulations enable us to compute
the fraction of the gas mass that resides, on average, within $r_{23.5}$. 
This fraction is found to be roughly 40 percent of the total gas mass. 
We are now fully equipped to compare observations and AM predictions.

\begin{figure*}
\includegraphics[width=0.47\textwidth]{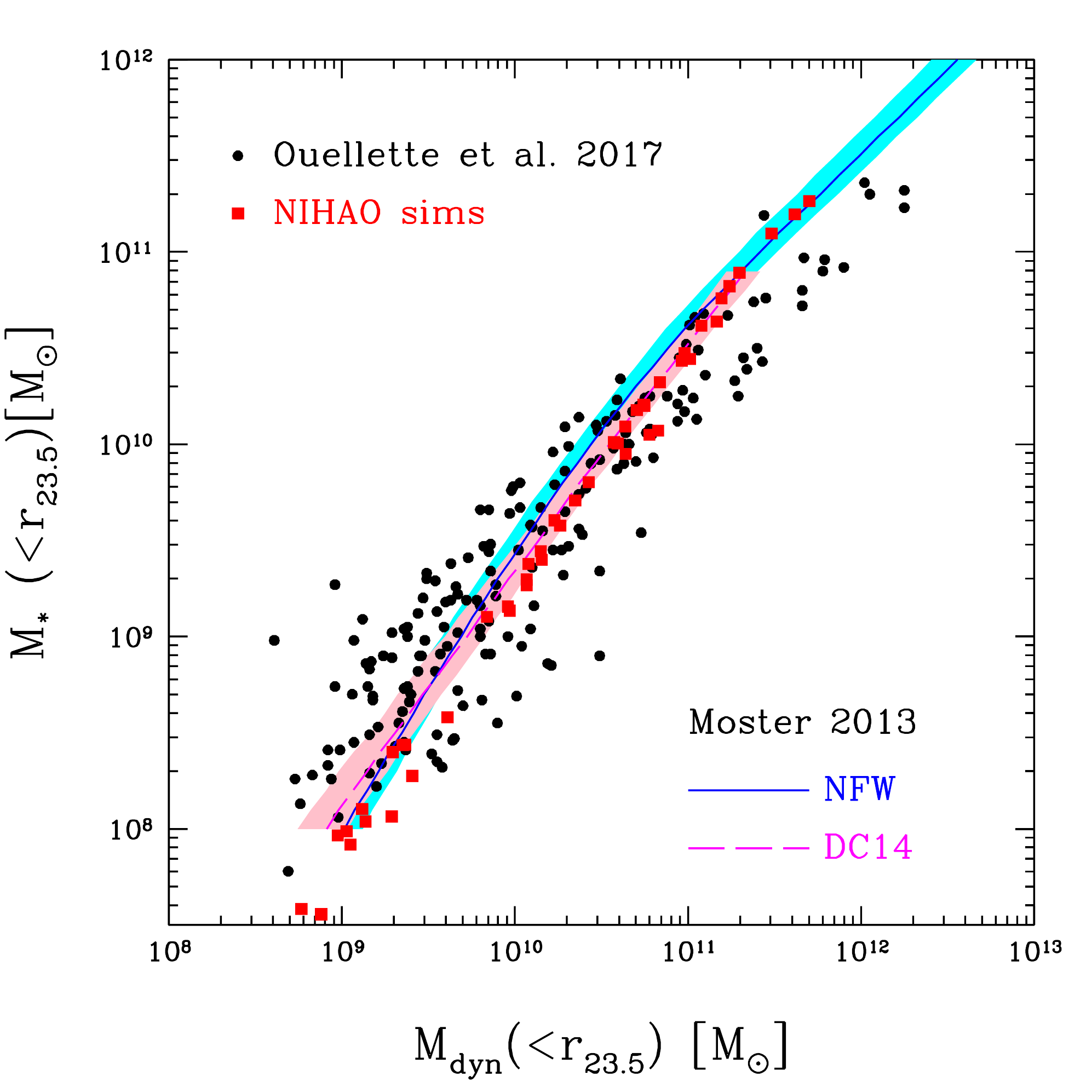}
\includegraphics[width=0.47\textwidth]{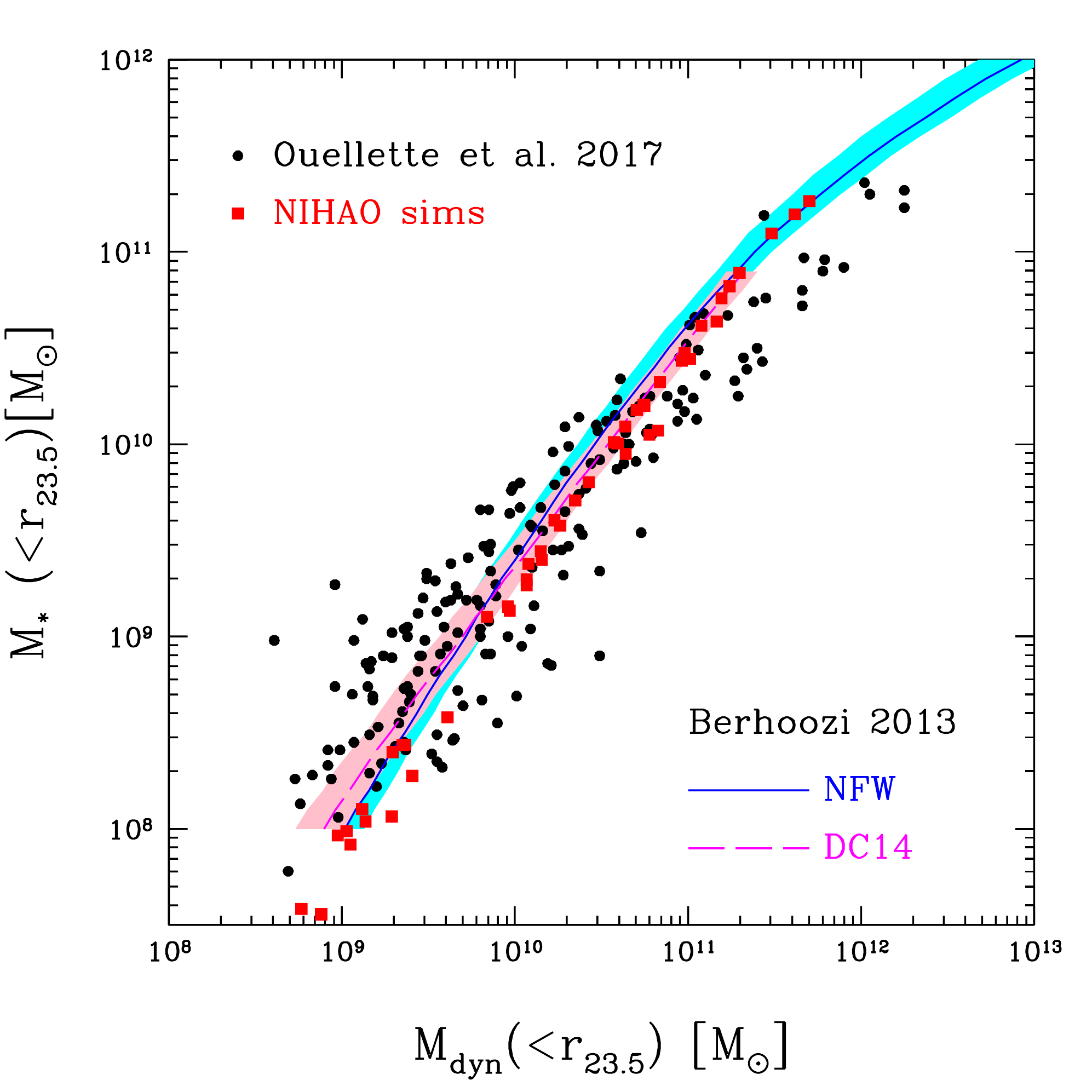}
\caption{Comparison of observations and AM predictions for the \citet[][left]{Moster2013}  and \citet[][right] {Behroozi2013}.
The blue solid line assumes an NFW profile for the DM distribution (this specific choice is inconsequential for our study), 
while the dashed magenta line is for the profile suggested by DC14. 
The corresponding shaded area represent the one-$\sigma$ scatter around the mean (see text for details).
The red squares represent the NIHAO simulations, while observations are shown as black points.}
\label{fig:AMmoster}
\end{figure*}

\Fig{fig:AMmoster} shows the final results for the SHMR proposed by \citet[][left panel]{Moster2013} and \citet[][right panel]{Behroozi2013}. 
The solid lines represent the results for two different choices for the dark matter profiles: NFW (blue), and DC14 (magenta).  
The black circles are the SHIVir data from \cite{Ouellette2017}.
The red squares show results for the NIHAO simulated galaxies. 
Results for the DC14 profile are only presented within the validity mass range of their fitting formula.

The shaded area in \Fig{fig:AMmoster} represents the scatter on the AM relation obtained from the procedure outlined above. 
Even though the (forward) scatter was computed for \Mdyn at a fixed \Mstar, we place \Mdyn on the abscissa axis as it is commoly presented in the literature.
We find that 80 percent of the scatter comes from the scatter in the (inverted) AM relation; the remainder of the scatter comes mainly from the concentration-mass relation \citep[0.11 dex,][]{Dutton2014}, the relation between stellar and cold gas masses \citep[0.32 dex,][]{Dutton2011}, and the relation between $r_{23.5}$ and \Mstar \citep[0.13 dex,][]{Ouellette2017,Stone2019}.

The data are well matched by the AM relation in the stellar mass range $10^8 - 5 \times 10^{10}~\Msun$,  and seem to favor the DC14 profiles rather than an unperturbed NFW.  NIHAO results closely reproduce the AM predictions based on the DC14 profile; this is expected since DC14 was based on an earlier version of the NIHAO simulations \citep{Stinson2013}.  Above a stellar mass of $\sim 5 \times 10^{10}~\Msun$, and regardless of the chosen dark matter profile, AM always over-predicts the stellar mass for a given dynamical mass.

Assuming that there is no bias in the  stellar mass profiles of NIHAO simulations \citep{Dutton2016}, 
this apparent mismatch may be resolved if the observed stellar masses are under-estimated at high mass.  This would indeed be expected in the context of stellar Initial Mass Function (IMF) variations between low to high mass systems. Such variations, suggesting a more top-heavy IMF at large stellar masses, are supported by different studies related to stellar populations  \citep[e.g][]{Conroy2012} and stellar internal dynamics \citep[e.g.][]{Cappellari2013, Dutton2013} among others.

At all scales, the predicted scatter in the AM relation is always significantly smaller than the observed one, indicating that the scatter in the data is dominated by observational errors. Finally our simulations have a smaller 
scatter than the one obtained from AM, consistent with results from other groups \citep[see][for a recent review on the galaxy-halo connection]{Wechsler2018}.



\section{Discussion and Conclusions}
\label{sec:conclusions}

The stellar-to-halo mass relation (SHMR), as predicted by Abundance Matching (AM) techniques, 
has become an essential benchmark for any model or simulation of galaxy formation.  Despite its
widespread use, the SMHR has only been tested directly against observations over a limited range
of masses and only for few galaxies \citep{Ouellette2017}.

In this paper, we have presented a comprehensive test of AM predictions based on 190 galaxies
of all morphological types, spanning over three orders of magnitude in stellar mass from $10^8$ to  $10^{11} $ \Msun.  
We have directly tested two of the most popular models for AM, namely the ones proposed by \citet{Behroozi2013} and \citet{Moster2013}. 

Instead of extrapolating observations to predict (total) stellar and viral  masses, we have recast
AM predictions in terms of stellar and total (dynamical) mass within the isophotal radius \r23_5 ;  
the latter quantities being readily accessible through observations. 
To properly translate AM theoretical predictions into tractable observables, we have used a mix of 
galaxy scaling relations and results from hydro simulations from the NIHAO suite \citep{Wang2015}.

We have found that AM predictions are a fair representation of observational data for stellar masses
below $\sim 5 \times 10^{10}$ \Msun, especially if we account for modifications of the dark matter
distribution due to galaxy formation as suggested by \citet{DiCintio2014a}. 

For a given dynamical mass, observations carry a scatter as large as one dex, much larger than
the intrinsic scatter in the AM relation, even when accounting for the scatter in the scaling relations
used to translate the original stellar-to-halo mass plane into the observational one.  The very small
predicted scatter in the AM relation is well reproduced by results from the  NIHAO simulated galaxies,
suggesting that most of the scatter in the observations comes from measurement uncertainties and possible environmental effects
\citep{Ouellette2017, Stone2019}.

Unfortunately, the large scatter in the observations thwarts any attempts to disentangle between
the two compared AM models.

For stellar masses above $\sim 5 \times 10^{10}$ \Msun, the AM models appear to overestimate
the stellar mass at a fixed dynamical mass with respect to observations.  Appealing to a top-heavy stellar Initial Mass Function \citep[e.g.][]{Conroy2012, Cappellari2013} at high masses might offer a possible solution.
More extensive coverage of the SHMR at high mass may also consolidate this issue. 


\section*{Acknowledgements}
The referee, Ben Moster, is warmly thanked for helping to improve the clarity of our paper. We also gratefully acknowledge the Gauss Centre for Supercomputing e.V. (www.gauss-centre.eu) 
for funding this project by providing computing time on the GCS Supercomputer SuperMUC at Leibniz
Supercomputing Centre (www.lrz.de) and the High Performance Computing resources at New York University Abu Dhabi. We are also grateful to the Natural Sciences and Engineering Research Council of Canada, the Ontario Government, and Queen's University for critical support through various scholarships and grants.

\newpage 



\bibliographystyle{mnras}
\bibliography{ref} 






\bsp	
\label{lastpage}
\end{document}